 \documentclass[preprint2]{aastex}

\usepackage{lscape}

\shorttitle{Speckle interferometry at SOAR in 2012-2013}
\shortauthors{Tokovinin, Mason, \& Hartkopf}
\begin{document}

% for float placement:
\renewcommand{\topfraction}{1.0}
\renewcommand{\bottomfraction}{1.0}
\renewcommand{\textfraction}{0.0}

\title{Speckle interferometry at SOAR in 2012 and 2013\altaffilmark{\dag} }

\altaffiltext{\dag}{Based on observations obtained  at the Southern Astrophysical Research
(SOAR) telescope,  which is a  joint project of the  Minist\'{e}rio da
Ci\^{e}ncia,  Tecnologia, e  Inova\c{c}\~{a}o (MCTI)  da Rep\'{u}blica
Federativa do Brasil, the  U.S. National Optical Astronomy Observatory
(NOAO), the  University of  North Carolina at  Chapel Hill  (UNC), and
Michigan State University (MSU).}

\author{Andrei Tokovinin}
\affil{Cerro Tololo Inter-American Observatory, Casilla 603, La Serena, Chile}
\email{atokovinin@ctio.noao.edu}

\author{Brian D. Mason \& William I. Hartkopf}
\affil{U.S. Naval Observatory, 3450 Massachusetts Ave., Washington, DC, USA}
\email{bdm@usno.navy.mil, wih@usno.navy.mil}

\begin{abstract}
We  report  the  results  of   speckle  runs  at  the  4.1-m  Southern
Astronomical Research (SOAR)  telescope in 2012 and 2013.   A total of
586 objects were  observed.  We give 699 measurements  of 487 resolved
binaries and upper detection  limits for 112 unresolved stars.  Eleven
pairs  (including  one  triple)  were  resolved for  the  first  time.
Orbital elements have been determined for the first time for 13 pairs;
orbits of another 45 binaries  are revised or updated.  
\end{abstract}


%-------------------------------------------------------------
\section{Introduction}

Knowledge of binary-star orbits is  of fundamental value to many areas
of astronomy.  They provide  direct measurements of stellar masses and
distances,  inform  us on  the  processes  of  star formation  through
statistics  of  orbital  elements,  and  allow  dynamical  studies  of
multiple stellar systems, circumstellar  matter, and planets.  A large
fraction of visual binaries are late-type stars within  100\,pc,
amenable to searches for  exo-planets. However, the current orbit catalog
contains some  poor or wrong  orbital solutions based  on insufficient
data.  To improve the situation,  we provide here new observations,
revise some orbits, and compute new ones.

This   paper  continues   the  series   of   speckle  interfero\-metry
observations    published    by   \citet[][hereafter    TMH10]{TMH10},
\citet{SAM09},  \citet{Hrt2012a}, and  \citet{Tok2012b}.  We  used the
same  equipment and  data  reduction methods.   All observations  were
obtained with  the 4.1-m SOAR  telescope located at Cerro  Pach\'on in
Chile.  Our  program is  focused on close  binaries with  fast orbital
motion,  where the  frequency of  measurements (rather  than  the time
span)  is critical for  orbit determination.   Some of  those binaries
were discovered  by visual observers, but most  are recent discoveries
made  by the {\em  Hipparcos} mission  and by  speckle interferometry,
including  our work at  SOAR. Spectroscopic  orbits are  available for
several fast nearby binaries  resolved here.  In addition, we measured
close  binaries with  known orbits  to  verify and  improve them  when
necessary, and wider pairs for calibration and quality control.

Data on binary-star measures and orbits are collected  in  the
Washington Double  Star Catalog, WDS  \citep{WDS}\footnote{See current
version   at   \url{http://ad.usno.navy.mil/wds/}}  and   associated
archives such as the {\em 4$^{\it th}$ Catalog of Interferometric 
Measurements of Binary Stars}, INT4
\citep{INT4}\footnote{\url{http://ad.usno.navy.mil/wds/int4.html}}
and the {\it 6$^{th}$ Orbit Catalog of Orbits of Visual Binary Stars}, VB6
\citep{VB6}.\footnote{\url{http://ad.usno.navy.mil/wds/orb6.html}}
These resources are extensively used here.

Section~\ref{sec:data}  recalls the  observing technique  and presents
new  measures,  discoveries,  and  non-resolutions.  New and updated 
orbits of 13 and 45 systems are given in Section~\ref{sec:orb-new} and
Section~\ref{sec:orb-rev}, respectively.

%======================================================================

%-------------------------------------------------------------
\section{New speckle measures}
\label{sec:data}

%-------------------------------------------------------------
\subsection{Instrument and observing method}

The   observations  reported   here  were   obtained  with   the  {\it
  high-resolution camera} (HRCam) -- a fast imager designed to work at
the  4.1-m SOAR  telescope  \citep{TC08}. For  practical reasons,  the
camera was  mounted on  the SOAR Adaptive  Module \citep[SAM,][]{SAM}.
However,  the laser guide  star of  SAM was  not used;  the deformable
mirror   of  SAM   was  passively   flattened  and   the   images  are
seeing-limited. The SAM module corrects for the atmospheric dispersion and
helps to  calibrate the  pixel  scale and orientation  of HRCam, as
explained below. The transmission curves of HRCam filters are given in
the                     instrument                    
manual\footnote{http://www.ctio.noao.edu/new/Telescopes/SOAR/Instru\-ments/SAM/archive/hrcaminst.pdf}.
We  used  mostly  the  Str\"omgren  $y$ filter  (543/22\,nm)  and  the
near-infrared $I$ (774\,nm) filter.

Observation of  an object  consists of accumulation  of 400  frames of
200x200 pixels each  with exposure time of 20\,ms  or shorter.  Frames
of 400x400 pixels were recorded  for pairs with separation larger than
1\farcs5.  Each object was normally recorded twice and these two image
cubes were processed independently.  Parameters of resolved binary and
triple  systems  are  determined  by  fitting a  model  to  the  power
spectrum, as explained in TMH10.

\subsection{Calibration of scale and orientation}

The  star light  reaches HRCam  after reflections  from  two non-rigid
mirrors -- the  thin active primary mirror of  SOAR and the deformable
mirror of  SAM.  Both can affect  the plate scale.   We calibrated the
transfer optics of  the SAM instrument by imaging  a single-mode fiber
located at the  telescope focus (mounted on the  SAM guide probe). The
fiber was translated by a  micrometer stage, allowing us to accurately
determine the  detector orientation  relative to the  instrument frame
and the pixel size in microns  projected from the detector to the SOAR
focal  plane.   The  pixel  size  at  focus  is  5.01\,$\mu$m,  stable
throughout the runs to better than 0.5\%.  Considering this stability,
we adopted a fixed pixel scale of 15.23\,mas/pixel as in the previous 
papers of this series.

The same  internal calibration  was applied to  the CCD  imager, SAMI,
which is  part of the  SAM Instrument. It has  4096$\times$4112 pixels
and  covers  a $3'$  square  field.  Thus,  both  HRCam  and SAMI  are
inter-calibrated.  Selected sky images taken with SAMI during the same
runs were corrected  for the optical distortion in  SAM and calibrated
against known positions of  stars using the {\em autoastrometry} tool\footnote{See 
\url{http://www.astro.caltech.edu/\~{}dperley/programs/auto\-astrometry.py}}
and the  2MASS catalog.  The pixel  scale of SAMI  and its orientation
were thus determined and then  translated to the orientation of HRCam.
We noticed that  the orientation of SAM was stable  during each run to 0\fdg1,
 but  changed between the  runs, although the  instrument was
not dismounted from the SOAR  Nasmyth rotator. Therefore, the angle of
the SOAR  rotator was not very  stable.  Moreover, it  should depend on
the  telescope orientation  if  the telescope  pointing  model is  not
perfect. With  this caveat  in mind, we  estimate the accuracy  of the
angle calibration to be about 0\fdg5. 

\begin{deluxetable}{l c c } 
\tabletypesize{\scriptsize}    
\tablecaption{Observing runs
\label{tab:runs} }                    
\tablewidth{0pt}     
\tablehead{ 
\colhead{Dates}  & 
\colhead{$\theta_0$} & 
\colhead{$N_{\rm obj}$} }
\startdata
Oct 29-30, 2012 & \phn2.10 & \phn19 \\ %sam12e
Dec 1-2, 2012   & \phn0.08 &    144 \\ %sam12f
Jan 28-29, 2013 & \phn1.50 & \phn11 \\ %sam13a
Feb 15-16, 2013 &  $-$1.00 &    268 \\ %sam13b
Mar 27, 2013    &  $-$0.21 & \phn41 \\ %sam13c
Jun 22, 2013    &  $-$0.21 & \phn38 \\ %sam13d
Sep 25-27, 2013 &  $-$0.66 &    174  %sam13e
\enddata
\end{deluxetable}

Table~\ref{tab:runs}  lists the  observing runs,  the  angular offsets
$\theta_0$ determined for each run, and the number of objects covered.
In February and  September 2013, we used the  telescope time allocated
for this program. The time allocation  in June 2013 was lost to clouds
almost   entirely.    The  remaining   data   were  collected   during
commissioning  runs of  the SAM  instrument, as  a backup  program. On
three  of those occasions,  we used  only few  available hours,  but a
whole night was spent in December 2012.

\subsection{Impact of telescope vibrations}

\begin{figure}[ht]
\epsscale{1.0}
\plotone{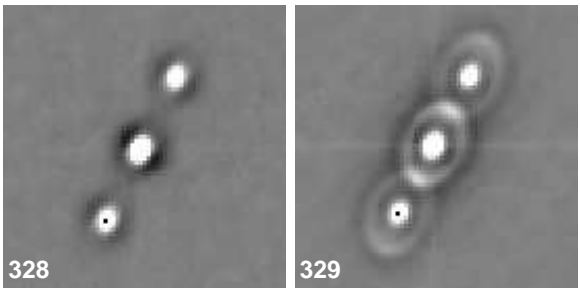}
\caption{Auto-correlation  functions  of the  binary  star HU~304  with
  separation  of  0\farcs288,  computed  from  image  cubes  \#328
  (exposure 5\,ms, left) and \#329 (exposure 20\,ms, right). The black dot marks 
the companion.
 \label{fig:vibr}}
\end{figure}

Previous work with HRCam,  SAM, and other SOAR instruments established
that the  telescope vibrates with the power-line  frequency of 50\,Hz and
the rms  amplitude reaching 40\,mas  in the worst cases.   The optical
axis  moves  on  an  elliptical  path  with  variable  elongation  and
amplitude;  the amplitude  increases  with the  zenith distance.   The
vibrations  are  not  stationary.  Attempts  to  locate  their  source
(e.g.  compressors of  the closed-cycle  coolers) have,  so  far, not
produced definitive results.

During the standard exposure time of 20\,ms, one full vibration period
is  sampled,  the  speckles   are  blurred,  and  the  resolution  and
sensitivity are  seriously degraded. Using a shorter  exposure time of
5\,ms  helps   to  recover   the  signal,  at   the  cost  of  reduced
flux. Figure~\ref{fig:vibr}  illustrates the effect  of shortening the
exposure  time by two observations  of  the same binary  star obtained
sequentially under  the same conditions. The elliptical  motion of the
SOAR's optical axis  produces characteristic ``disks'' which sometimes
blur the speckle completely.

We used the  exposure time of 5\,ms and went as short as 2\,ms when possible.
Fainter stars were observed with the longer exposure of 20\,ms, and in
this case  the success depended on  the presence and  amplitude of the
vibrations.  The February 2013 run was particularly affected, while in
September 2013 the vibrations were less intense.

The faintest resolved binary is HIP~48273B = RAO~90 (WDS J09505+0421) at
$V=12.1$.  It was observed in the $I$ filter with 20-ms exposure time;
the  signal is  affected by  vibrations.   The next  faintest pair  is
KUI~41 (WDS J09310$-$1329) at $V=10.7$, observed again in $I$, but with
a 5-ms exposure.  With a  normal 20-ms exposure, stars $1.5^m$ fainter
could be  recorded. The  2$\times$2 binning of  the detector  helps to
increase  the sensitivity  for  faint stars  while under-sampling  the
speckle, but it was not used here.

\subsection{Data tables}

Table~2  lists   699  measures  of 487  resolved   binary  stars  and
sub-systems, including newly resolved pairs. The format did not change
from the previous papers.  The  columns of Table~2 contain (1) the WDS
designation, (2) the ``discoverer designation'' as adopted in the WDS,
(3) an alternative name, mostly  from the {\it Hipparcos} catalog, (4)
Besselian epoch  of observation, (5) filter, (6)  number of individual
data  cubes, (7,8)  position angle  $\theta$ in  degrees  and internal
measurement  error in tangential  direction $\rho  \sigma_{\theta}$ in
mas,  (9,10) separation $\rho$  in arcseconds  and its  internal error
$\sigma_{\rho}$ in  mas, and (11) magnitude difference  $\Delta m$. An
asterisk follows   if $\Delta m$  and the true  quadrant are
determined from  the resolved  long-exposure image; a  colon indicates
that the data  are noisy and $\Delta m$  is likely over-estimated (see
TMH10 for  details).  Note  that in the  cases of multiple  stars, the
positions  and photometry  refer  to the  pairings between  individual
stars, not with photo-centers of sub-systems.

For  stars with known  orbital elements,  columns (12--14)  of Table~2
list  the residuals to  the ephemeris  position and  the 
references to the orbits. 

%code  of   reference  to   the  orbit  adopted   in  VB6.\footnote{See
%  \url{http://ad.usno.navy.mil/wds/orb6/wdsref.html}} 

We did  not use image  reconstruction and measure the  position angles
modulo $180^\circ$.  Plausible quadrants  are assigned on the basis of
orbits or prior  observations, but they can be  changed if required by
orbit calculation. For triple  stars, however, {\em both} quadrants of
inner and  outer binaries have  to be changed  simultaneously; usually
the  slowly-moving  outer  pair  defines  the quandant  of  the  inner
sub-system without ambiguity.

Our  software does  not have  capability  of fitting  models of  power
spectra with more than  three components.  Sub-systems in the resolved
quadruple    stars   RST~244Ba,Bb   (WDS J07185$-$5721)    and   MLO~3Ba,Bb
(WDS J13147$-$6335)  were measured  crudely using the  peaks  in the
autocorrelation functions.

Table~3 contains  the data on 112  unresolved stars, some  of which are
listed   as  binaries   in  the   WDS  or   resolved  here   in  other
filters. Columns (1) through (6)  are the same as in Table~2, although
Column (2)  also includes other  names for objects  without discoverer
designations.  Columns  (7,8) give  the  $5  \sigma$ detection  limits
$\Delta m_5$  at $0\farcs15$ and  $1''$ separations determined  by the
procedure  described  in TMH10.   When  two  or  more data  cubes  are
processed, the best detection limits are listed.

\subsection{New pairs}
\label{sec:new}

For the reader's convenience, we extracted data on the binaries resolved
here for  the first time into  Table~4. There are 11  objects, some of
which are newly resolved visual triples. As in the previous runs, we
tried to observe nearby solar-type  dwarfs that are known to be binary
by  their variable  radial  velocity  (RV), mostly  from  the work  by
\citet{N04},  or by astrometric acceleration  \citep{MK05}.  This
continues the work on resolving  astrometric binaries that was done at
SOAR  and at Gemini  \citep{astrom1,astrom2}.  Comments  on individual
stars follow.

{\bf 02321$-$1515 = HIP~11783} is an F5V dwarf at 27\,pc. It has variable
RV and  astrometric acceleration.  The companion was  resolved only in
the $I$ band.  The star is bright, $V=4.74^m$; it was observed through
clouds when no other program objects could be accessed.  There is also
a common proper motion companion  B = HIP~11579 of spectral type K2.5V
located at $345''$  and 252\fdg5 from A \citep{LEP},  making the whole
system  triple.   The proper  motions  of  A  and B  differ  slightly,
possibly because of the inner system Aa,Ab with an estimated period of
$\sim$10\,yr.

{\bf 04007+2023 =  HIP~18719} is a spectroscopic binary  in the Hyades
with   period  16.7\,yr,  estimated   semi-major  axis   of  0\farcs12
\citep{Griffin2012},  and  astrometric  acceleration.   The  secondary
lines were  detected by \citet{Bender2008}, who evaluated  the mass of
the  secondary component  at 0.4${\cal{M}}_{\odot}$.   The  object was
tentatively  resolved  with   Robo-AO  \citep{RoboAO}  in  2012.757  at
$342^\circ$, 0\farcs142 and $\Delta i  = 2.8$.  The resolution here is
secure, at smaller  separation and a very different  position angle of
$96^\circ$ (or $276^\circ$) with $\Delta I = 3.6$.

{\bf  07435+0329  =  HIP~37645}  has  a  variable  RV.   According  to
D.~Latham (private communication), it  is a single-lined binary with a
period  on the  order of  35\,yr.  Such  a period  corresponds  to a
semi-major  axis of  0\farcs27,  of  the same  order  as the  measured
separation.   This is  a triple  system with  physical companion  B at
9\farcs6  (AB is STF~1134,  discovered in  1832).  The  more distant
companion C listed  in the WDS is optical, as  evidenced by the motion
of AC since 1906.  Although the system AC is designated as STF~1134AC,
it was actually first measured in 1906 by Burnham (1913).

{\bf 08391$-$5557 = HIP 42424}  is the known binary HU~1443, measured here
at 0\farcs917  separation. Its secondary  component turns out to  be a
new pair Ba,Bb with 0\farcs07 separation. This example shows the power
of speckle  interferometry at a 4-m  telescope. The outer  pair AB was
observed with  speckle interferomety by E.~Horch et al. (2006) at smaller
telescopes  that are  sufficient  for its  resolution,  but the  close
sub-system escaped detection until now.

{\bf 09086$-$2550 = HIP~44874} is  similar to the previous object: the
known pair RST~2610 at 1\farcs78  hosts the new close 0\farcs12 binary
Ba,Bb.   The A  component is  itself a  close pair  with variable  RV and
astrometric acceleration.   It was already observed at  SOAR and found
unresolved (also unresolved here).   The orbital period and mass ratio
of the sub-system Aa,Ab thus remain indeterminate so far. In contrast,
the new  sub-system Ba,Bb  has an estimated  period of  $\sim$15\,yr and
component  masses of  0.6 and  0.5 ${\cal{M}}_{\odot}$  estimated from
their luminosity. The whole system is therefore a 2+2 quadruple.

{\bf   09383$+$0150  =   TOK~358}  was   also  resolved   with  Robo-AO
\citep{RoboAO} in  2013.0526 at 345\fdg1,  0\farcs447 and $\Delta  i =
2.34$.

{\bf  11192$-$1950 = RV  Crt  =  HD  98412} is  an  eclipsing  binary  of
$\beta$~Lyr  type, spectral type  F8.  The  existence of  the tertiary
companion was suggested from  the eclipse timing by T.~Armond (private
communication). The  object was observed  on her request  and resolved
into a 0\farcs05 pair of nearly equal stars.
 
{\bf 11420$-$1701  = HIP~57078} unexpectedly  turned out to be  a triple
star.  This is an F5V dwarf with astrometric acceleration \citep{MK05},
presumably produced by the newly resolved 0\farcs14 pair Aa,Ab with an
estimated  period of  $\sim$20\,yr.   In addition,  we  see a  distant
companion  B at  1\farcs11.  The  physical nature  of B  is yet  to be
confirmed  by  repeated measurement  within  a  year,  but is  likely,
considering the low density of background stars around this target.

{\bf 12509$-$5743 = HIP~62699}  is another nearby dwarf with astrometric
acceleration and  variable RV  resolved here at  0\farcs19. The estimated
period is $\sim 25$\,yr.

\begin{figure}[ht]
\epsscale{1.0}
\plotone{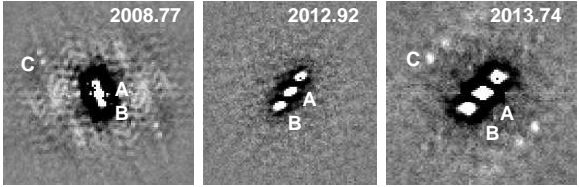}
\caption{Auto-correlation  functions  of  the resolved  triple  system
  VOU~35 observed  on three epochs at  SOAR. North is up,  East to the
  left, the scale is approximate. 
 \label{fig:VOU35}}
\end{figure}

{\bf 21368$-$3043 =  VOU 35~AB = HIP~106701} is a  G5V dwarf with {\em
  Hipparcos}  parallax of  16.1\,mas. The  projected separation  of AB
indicates a probable orbital period of about 20\,yr (see the orbit with
$P=19.8$\,yr in Table~6, Figure~\ref{fig:14}). We detect another faint  
component C at 0\farcs45 and 53\fdg5 from A. Re-examination of archival 
speckle data show that the component C was not seen on 2008.54 (it was 
fainter than the detection limit), but was    present on 2008.7669  at 
approximately 62\fdg9  and  0\farcs475 (in  the  $y$ filter). However,  
the next observation on 2012.923 with good signal-to-noise ratio shows 
no trace of the C component,  while  it  is  securely   detected   now
(Figure~\ref{fig:VOU35}).

The multiple system HIP~106701  has been detected in X-rays \citep[see
  also][]{SACY}.  The disappearance of C in 2012.92 could be caused by
its variability  (e.g. eclipses).  The  projected separation of  AC is
$\sim$30\,AU  and  implies a  period  on  the  order of  100\,yr.  The
observed motion of  AC (10$^\circ$ in 5\,yr) does  not contradict this
estimate. If C were some  unrelated background star, the proper motion
of  A   (0\farcs11  per  yr)   would  have  drastically   changed  the
configuration of AC in 5\,yrs.

Interestingly, there is another star in the 2MASS catalog at 234\fdg0,
11\farcs2  from   A.   Its  infrared  magnitudes  match   a  dwarf  of
$\sim$0.65\,${\cal{M}}_{\odot}$ at the  same distance. This additional
component D is  visible at similar relative position  in the saturated
DSS image. Considering also low crowding, we tentatively conclude that
the  pair  AD is  physical  and  that the  whole  system  is at  least
quadruple.    For    completeness,   we   list   the    AD   pair   in
Table~\ref{tab:new}. 

%{\bf ** 2MASS and UCAC measures extracted - should we include both in the table?}

\subsection{Comments on individual objects of interest}
\label{sec:cmt}

{\bf Potentially  spurious binaries.}  A binary system  can become too
close and unresolvable when it goes through periastron. It is expected
to  re-appear after a  few years  if the  estimated orbital  period is
short. Repeated  observations of several promising  candidates at SOAR
failed to resolve them,  however, despite much improved resolution and
dynamic range of speckle  interferometry in comparison with the visual
observations used  to discover  these binaries. Table~5  lists several
such  cases,   adding  to  the   list  of  mysterious   ``ghosts''  in
\citet{Tok2012b}.  It gives the  year of last measurement according to
the WDS, the period of  speckle non-resolution, and the number $N_{\rm
  UR}$ of negative speckle  observations.  The orbit with 21-yr period
computed for B~594 by  \citet{Nrr1983} predicts its separation between
0\farcs07 and 0\farcs12 during the period of non-resolutions at SOAR.

{\bf Reversed  quadrant of  the orbit} is  evidenced in  some resolved
triple  systems where  the orientation  of  the inner  (fast) pair  is
determined by the outer (slow) binary. In these cases, reversal of the
observed  quadrant (allowed  in  classical speckle)  is not  possible;
instead, the orbital node must be changed by $180^\circ$. Such triples
here are FIN~337BC (WDS J01198$-$0030), FIN~308AB (WDS J10282$-$2548),
and RBT~1Aa,Ab (WDS J14038$-$6022).

{\bf  02426$-$7947 =  HIP~12654}  is the  acceleration binary  TOK~362
resolved  with NICI  \citep{astrom2}.   The new  measure confirms  the
hypothesis that the pair opens  up.  Note that the secondary component
is red:  $\Delta V=4.2$, $\Delta  I = 2.6$,  while $\Delta K  = 0.61$.
The secondary  could itself  be a close  pair of M-dwarfs.   The large
$\Delta V$ and presumably close  separation in 1991 help to understand
why this pair was not resolved by {\em Hipparcos}.

{\bf 04311$-$4522 = HIP 21079} is the acceleration binary TOK~208
resolved in \citet{astrom1}. The wide separation implies a long 
estimated orbital period of $\sim$600\,yr, making it difficult to
explain the acceleration, unless the system is triple.

{\bf 06454$-$3148 = HIP~32366}. We discovered independently the close
pair  Ba,Bb, converting  this nearby  solar-type dwarf  into  a triple
system.  In  fact, this system was resolved  earlier by \citet{EHR10}.
They  even suggested  a preliminary  circular  orbit of  Ba,Bb with  a
period  of   3.5\,yr,  based  on  several   measurements.   The  small
astrometric acceleration of A could be caused by its attraction to 
component B.

{\bf 06573$-$4929 = RST~5253~AC.}  The wide companion C was discovered
at SOAR on  2010.97 at 1\farcs05, 236\fdg9; the  quadrant published in
\citet{Hrt2012a} was  chosen wrongly.  It  is measured here  at 1\farcs06
and  238\fdg6. The  quadrant is now confirmed from the direct image and
matches the orbit of the inner pair AB.

{\bf 09383+0150 = HIP~47292}  is an acceleration binary, also resolved
in 2013.05 with Robo-AO (Riddle et al.\ 2014).

{\bf 14038$-$6022  = VOU 31 and  RBT 1 Aa,Ab} is the spectacular triple
system $\beta$~Cep.

%---------------------------------------------------------------------
\section{First orbits}
\label{sec:orb-new}

In  this  Section,  we  derive  first  orbits  for  some  binaries  or
sub-systems observed here. This continues the work of \citet{Hrt2012a}
on  cleaning and improving  the VB6  catalog. We  refer to  that paper
where  the  method of  orbit  calculation  is  described. The  orbital
elements of  13 pairs are listed  in Table~6. As  a consistency check,
the mass  sum resulting  from the new  orbits and the  {\em Hipparcos}
parallax is  given. We  discuss briefly some  of these objects  in the
remainder of this Section. Figures  3--15 show the new orbits. In each
of  those  figures, speckle  and  other  high-resolution measures  are
plotted  as filled  circles, the  measures  from this  work as  filled
stars, micrometric observations as plus signs, and the {\em Hipparcos}
observations as filled  diamonds. A line connects each  measure to its
predicted position on the orbital  ellipse. The dot-dashed line passing
through  the primary indicates  orbital nodes,  the grey  circle shows
diffraction limit of the 4.1-m telescope, with the scale in arcseconds
shown on both axes.  The orientation and  the direction of  motion are
indicated in the lower right corner of each figure.

{\bf  02022$-$2402 =  TOK~41~Ba,Bb} is  the secondary  sub-system with
nearly equal components in the triple star HIP~9497.  The 138-yr orbit
of the outer pair AB=HDS~272 \citep{Hrt2011d} is still preliminary and
corresponds to a mass sum of 11.4\,${\cal M}_\odot$. The 14-yr orbit 
of Ba,Bb  is not  yet fully  covered  and   also preliminary.  Further
speckle  monitoring  will lead  to the reliable orbits that will allow
dynamical  analysis  of  this  interesting  triple  system  with  weak
hierarchy, i.e. a small ratio of outer and inner periods.

\begin{figure}[ht]
\epsscale{1.0}
\plotone{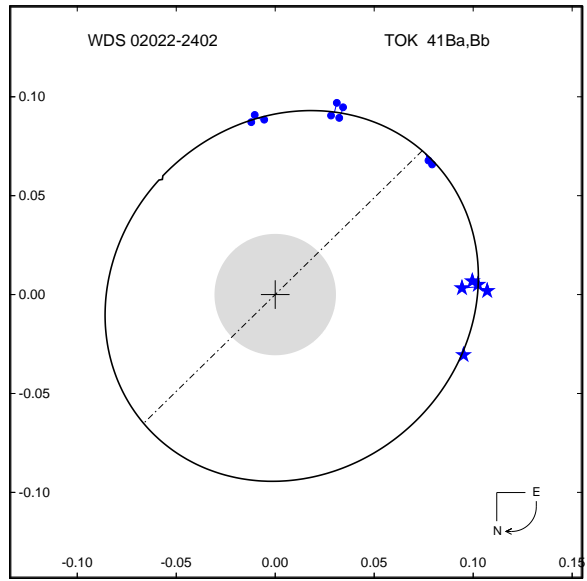}
\caption{Orbit of TOK~41~Ba,Bb. 
 \label{fig:3}}
\end{figure}

{\bf 02572$-$2458 = BEU~4~Ca,Cb}  is a spectroscopic pair belonging to
the quadruple system ADS~2442, known also as GJ~120.1.  The components
AB  are  HD~18455 =  HIP~13772  =  BU~741  (also measured  here),  the
component  C  is HD~18445  =  HIP~13769,  K2V.   It was announced as a
spectroscopic binary by  Duquennoy \& Mayor in 1991. Some elements of
its spectroscopic orbit  were published by \citet{Halbwachs2000}; they
mention orbit publication in a ``forthcoming paper'' which has not yet
appeared.  The low velocity  amplitude hinted at a secondary component
of planetary or brown-dwarf mass.  However, the system was resolved by
\citet{Beuzit2004}  in  2000.6.   The   large  mass  of  Cb  was  also
established   by   \citet{Reffert2011}   from  the   {\em   Hipparcos}
astrometry.  We  see now  that the orbit  is oriented  almost face-on,
explaining the low RV amplitude  (which is further reduced by line
blending).  In  orbit fitting, we  fixed some elements known  from the
spectroscopy and the poorly-constrained inclination.

\begin{figure}[ht]
\epsscale{1.0}
\plotone{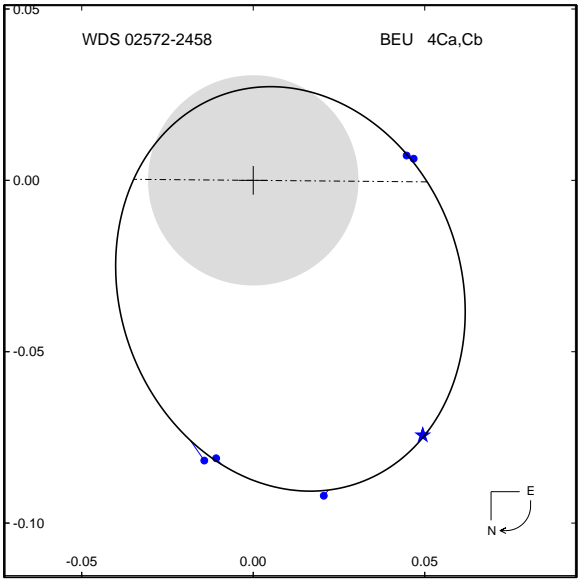}
\caption{Orbit of BEU~4~Ca,Cb.
 \label{fig:4}}
\end{figure}

\begin{figure}[ht]
\epsscale{1.0}
\plotone{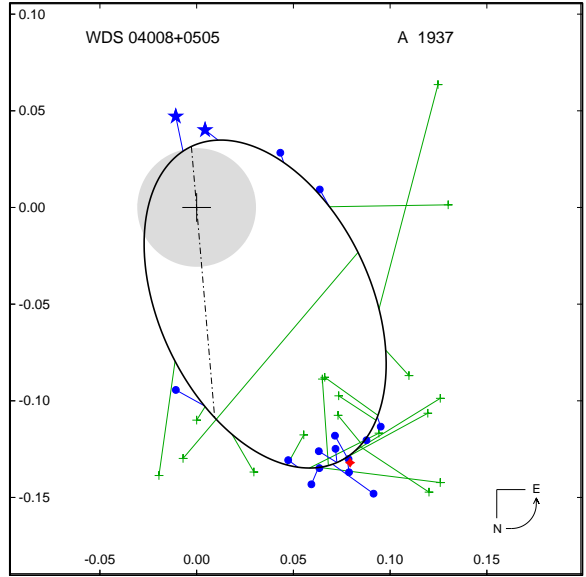}
\caption{Orbit of A~1937 
 \label{fig:5}}
\end{figure}

\begin{figure}[ht]
\epsscale{1.0}
\plotone{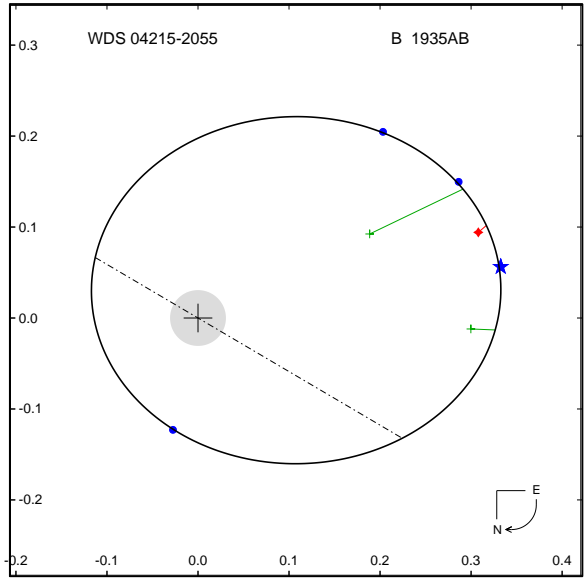}
\caption{Orbit of B~1935
 \label{fig:6}}
\end{figure}

{\bf 05072$-$1924 = FIN~376 =  HIP~23818 = HD 33095} is a double-lined
spectroscopic binary  with a period of 3.9\,yr  according to D.~Latham
(private communication). The short period explains why no visual orbit
has been  computed so far,  as the orbital coverage  was insufficient.
The orbit in Table~6 is  derived by combining radial velocities of the
two  components kindly  made available  to  us by  D.~Latham with  the
speckle and visual measurements, which explains the  small errors.  We
do  not publish  here  the spectroscopic  elements  and the  resulting
masses, deferring this analysis to a future paper.

\begin{figure}[ht]
\epsscale{1.0}
\plotone{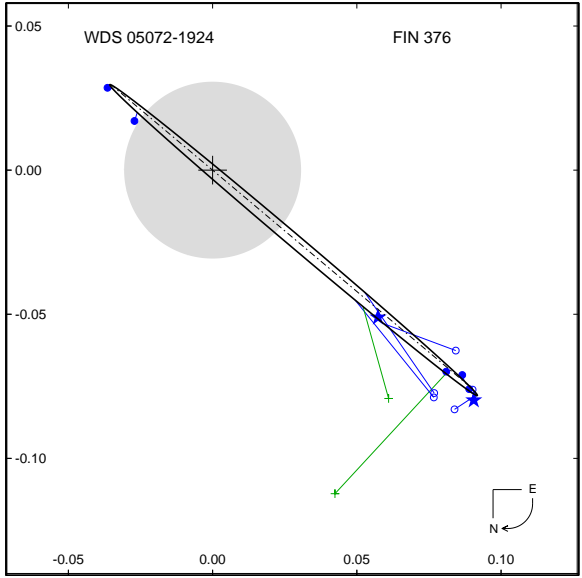}
\caption{Orbit of FIN~376.
 \label{fig:7}}
\end{figure}

\begin{figure}[ht]
\epsscale{1.0}
\plotone{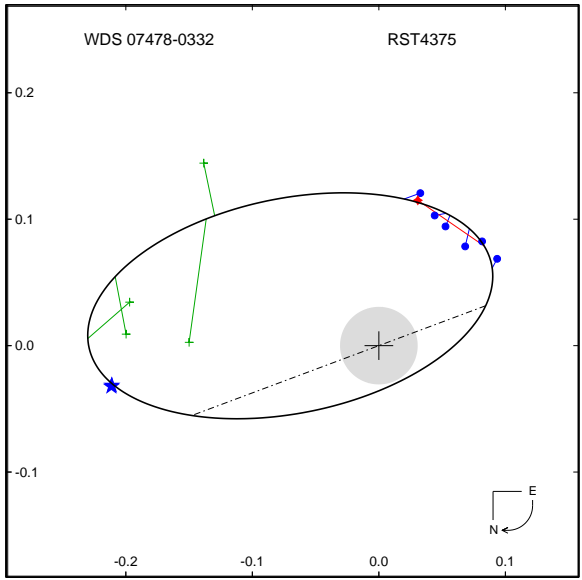}
\caption{Orbit of RST~4375
 \label{fig:8}}
\end{figure}

{\bf 07478$-$0332 = RST~4375  = HIP~38039.}  The speckle measures from
the 1990s appear to be located  on a straight line, rather than on an
ellipse (Figure~\ref{fig:8}).   The orbit is  still very preliminary.
The mass sum of 6.5\,${\cal{M}}_{\odot}$ seems a bit too large for the
spectral  type  A0.   Considering  the  large deviation  of  the  {\em
  Hipparcos}  measurement  from the  speckle  data,  we might  suspect
inaccurcy  of its  parallax.

{\bf 09191$-$4128  = CHR~239 =  HIP~45705.}  This first  orbit appears
quite reliable already and gives  a reasonable mass sum (spectral type
G2V).  This is also an acceleration binary in {\em Hipparcos}.
%mag. diff 2.2 (y) and 1.6 (I).

\begin{figure}[ht]
\epsscale{1.0}
\plotone{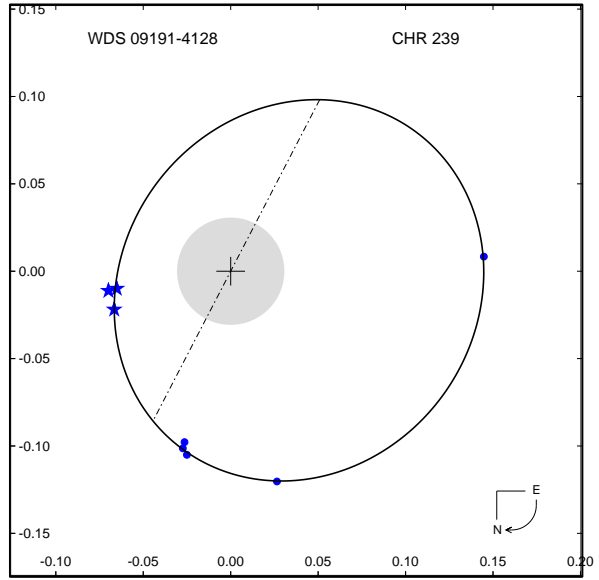}
\caption{Orbit of CHR~239
 \label{fig:9}}
\end{figure}

{\bf 11514$+$1148 = HDS~1672 = HIP~57821.}  The 54-yr orbit is not yet
fully covered and remains  preliminary.  The mass sum matches spectral
type  F6V.  It  became  possible to  compute  the orbit  owing to  the
observations at SOAR and by \citet{Horch02,Horch08}.

\begin{figure}[ht]
\epsscale{1.0}
\plotone{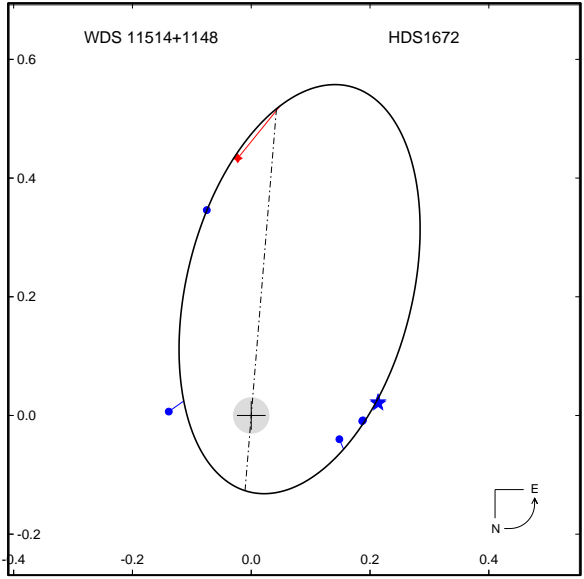}
\caption{Orbit of HDS 1672
 \label{fig:10}}
\end{figure}

{\bf 11525$-$1408 = HDS~1676  = HIP~57894.}  This binary has completed
more  than   one  revolution  since  its  first   resolution  by  {\em
  Hipparcos.}   The  coverage remains  poor,  however (exclusively  at
SOAR).  The mass sum matches spectral type G0V.

\begin{figure}[ht]
\epsscale{1.0}
\plotone{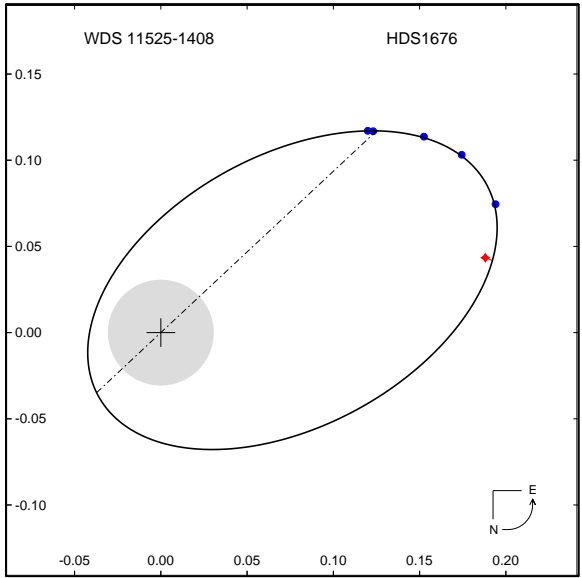}
\caption{Orbit of HDS 1676
 \label{fig:11}}
\end{figure}

{\bf 12485$-$1543 = WSI~74~Aa,Ab =  HIP~62505 = HD~111312 } is a K2.5V
star GJ~1165 within 25\,pc  of the Sun.  \citet{Raghavan10} state that
it is a double-lined spectroscopic  binary with a period of 2.698\,yr,
but  do not  give its  orbital  elements.  Radial  velocities of  both
components kindly provided by  D.~Latham were included in the combined
orbital  solution. We publish  here only  the ``visual''  elements and
defer further analysis to another paper, as in the case of FIN~376.

\begin{figure}[ht]
\epsscale{1.0}
\plotone{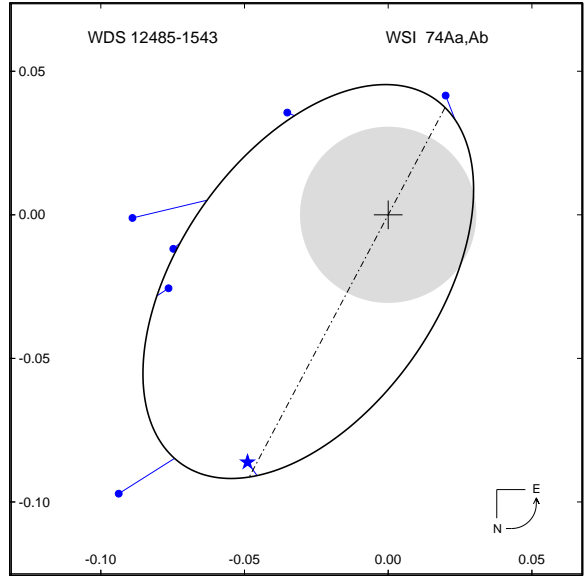}
\caption{Orbit   of   WSI~74   Aa,Ab.   
 \label{fig:12}}
\end{figure}

{\bf 17066+0039 = TOK~52~Ba,Bb =  HIP~83716} was discovered at SOAR in
2009 (cf. TMH10), its first orbit  is computed here with almost a full
revolution covered.  The outer pair BU~823AB also has a computed orbit
with a period  of 532\,yr, which may be nearly  co-planar with that of
Ba,Bb.   The  mass sum  of  Ba,Bb  estimated  from its  luminosity  is
1.5\,${\cal{M}}_{\odot}$ and matches  well the dynamical mass inferred
from the orbit.

\begin{figure}[ht]
\epsscale{1.0}
\plotone{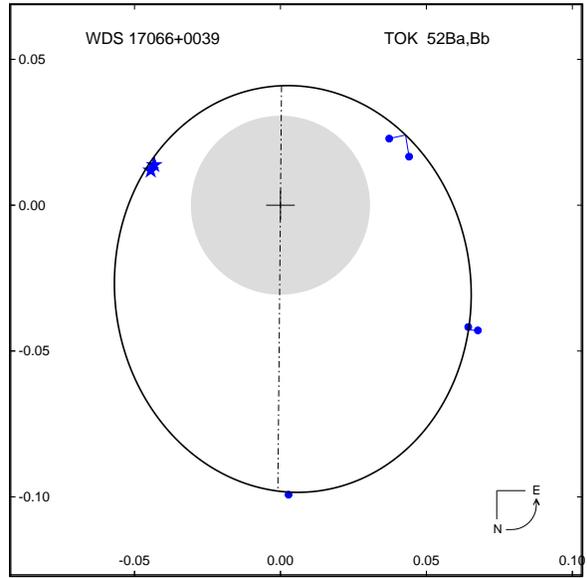}
\caption{Orbit of TOK 52 Ba,Bb
 \label{fig:13}}
\end{figure}

\begin{figure}[ht]
\epsscale{1.0}
\plotone{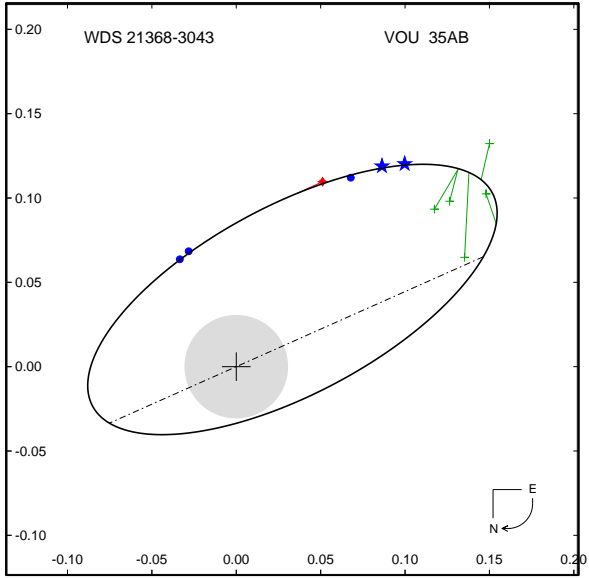}
\caption{Orbit of VOU~35AB.
 \label{fig:14}}
\end{figure}

\begin{figure}[ht]
\epsscale{1.0}
\plotone{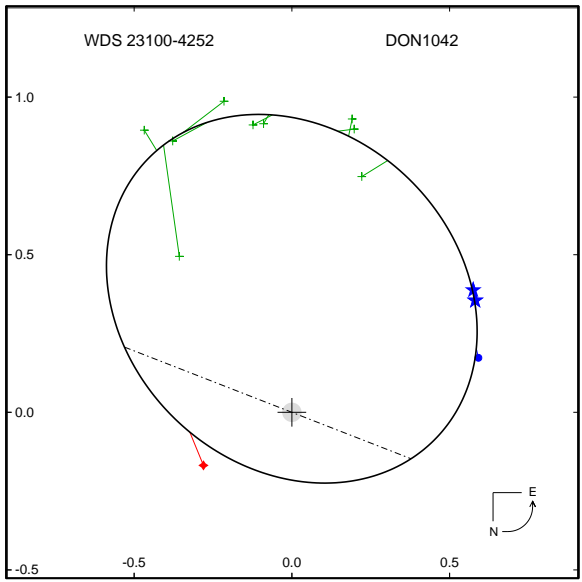}
\caption{Orbit of DON 1042.
 \label{fig:15}}
\end{figure}

%---------------------------------------------------------------------
\section{Revised orbits}
\label{sec:orb-rev}

\begin{figure}[ht]
\epsscale{1.0}
\plotone{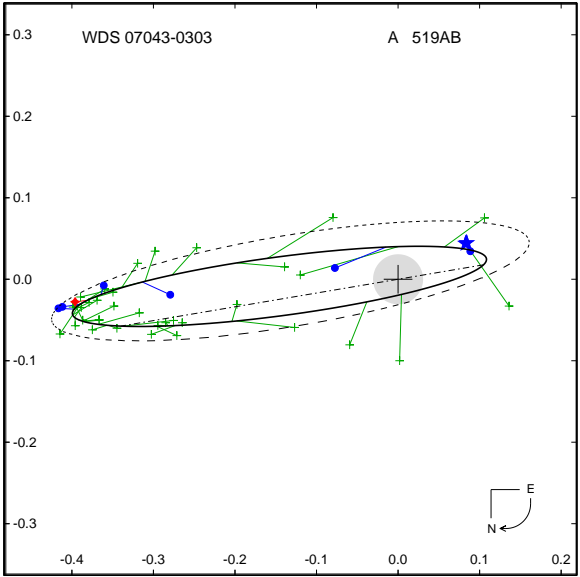}
\caption{Minor   revision of  the orbit of  A~519 (WDS J07043$-$0303),
  illustrating the change of scale in comparison with the previous orbit by
  \citet{Doc2009g} (dashed line).
 \label{fig:r1}}
\end{figure}

Some  pairs show  substantial  deviations from  the published  orbits,
prompting their  revisions. In  several instances those  revisions are
only minor, as the orbit  was already well constrained by the existing
data. These revisions simply reduce  errors of the elements or correct
systematics such as scale of the orbit, see e.g.  Figure~\ref{fig:r1}.

\begin{figure}[ht]
\epsscale{1.0}
\plotone{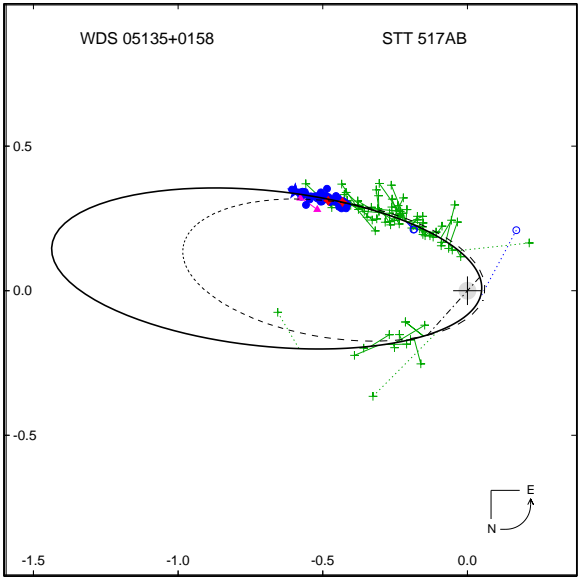}
\caption{Orbit of STT 517 AB, $P=992$\,yr. The actual period 
can be much longer (few thousand years), it is not yet constrained by
the observed arc. 
 \label{fig:r2}}
\end{figure}

The orbits of  many long-period binaries are not  yet fully covered by
the  observations; decades or  even centuries  of additional  data are
needed to do  so. Here, the revision just  improves the description of
the orbital  motion observed so  far, while the period  and semi-major
axis   remain    essentially   unconstrained,   as    illustrated   in
Figure~\ref{fig:r2}. These should provide reliable ephemerides over 
the next several decades.

\begin{figure}[ht]
\epsscale{1.0}
\plotone{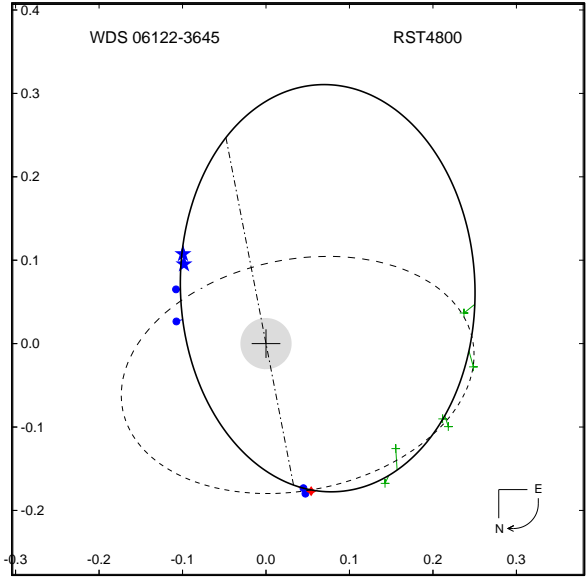}
\caption{Orbit  of RST  4800,  $P=174$\,yr. New  observations at  SOAR
  cause drastic revision of the previous orbit by \citet{USN2002}. The new orbit
  is  still lacking  coverage and  will  be further  corrected in  the
  coming decades.
 \label{fig:r3}}
\end{figure}

Substantial  or   drastic  revisions   of  existing  orbits   are  not
uncommon. This happens when   orbits were computed prematurely with
insufficient  or inaccurate  observations. In  such cases,  the revised
orbit had to  be calculated from scratch. Hopefully,  these new orbits
are closer to the true ones and will be corrected incrementally in the
future. The mass sum computed from the new orbits and known parallaxes
is  not substantially different  from its  estimate based  on spectral
type or luminosity of the stars. See Figure~\ref{fig:r3}.

Table~7 lists elements of 45 orbits revised here. For orbits of grades
1--3,  the  errors  of  each   element  are  given,  while  the  still
preliminary orbits of grades 4 and 5 are given without formal errors.
The last column of Table~7 contains references to the previous orbital
solutions. Considering the availability of orbital plots in the VB6
on-line catalog, we do not provide them here, except three
illustrative cases in Figures~\ref{fig:r1}--\ref{fig:r3}. 

The 25-yr  astrometric sub-system belonging to the  A-component of the
bright  visual  binary   $\zeta$~Aqr  (WDS  J22288$-$0001)  was  first
resolved at  SOAR and is  listed in the  WDS as EBE~1.  Its  orbit is
still  very preliminary,  with  period and  eccentricity fixed.   This
mutiple system was recently discussed by \citet{Hrt2012a}.

\acknowledgments   We  thank  the   operators  of   SOAR  D.~Maturana,
S.~Pizarro, P.~Ugarte, A.~Past\'en for their help with labor-intensive
speckle observations, and the anonymous referee for careful reading of
the manuscript and tables.  This work used the SIMBAD service operated
by Centre des Donn\'ees Stellaires (Strasbourg, France), bibliographic
references from  the Astrophysics Data System  maintained by SAO/NASA,
and the Washington Double Star Catalog maintained at USNO.
%

%-------------------------------------------------------------
%-------------------------------------------------------------
%-------------------------------------------------------------
%-------------------------------------------------------------
%-------------------------------------------------------------

{\it Facilities:}  \facility{SOAR}.

%{\bf Table 2: minor corrections
%02022$-$2402: negative $\Delta$m doesn't line up
%07185$-$5721: add phn after separation for Ba,Bb pair
%22288$-$0001: change DD back to EBE 1 to agree with WDS, Table 7, and text.}

%\clearpage

%\begin{landscape}
%\input{measures.tex}
%\end{landscape}
%\input{single-table.tex}

%\clearpage

%\begin{landscape}

\begin{flushleft}

\begin{deluxetable}{l l l  ccc  rc cc l r r l }                                                                                                                                
\tabletypesize{\tiny}                                                                                                                                                                                                                                                                                                                             
%\tabletypesize{\scriptsize}                                                                                                                                                                                                                                                                                                                             
\tablecaption{Measurements of binary stars at SOAR                                                                                                                            
\label{tab:double} }                                                                                                                                                            
\rotate
%\tablewidth{500pt} 
\tablewidth{0pt} 
\tablenum{2}                                                                                                                                                         
\tablehead{                                                                                                                                                                     
\colhead{WDS} & \colhead{Discoverer} & \colhead{Other} & \colhead{Epoch} & \colhead{Filt} & \colhead{N} & \colhead{$\theta$} & \colhead{$\rho \sigma_{\theta}$} &               
\colhead{$\rho$} & \colhead{$\sigma \rho$} & \colhead{$\Delta m$} & \colhead{[O$-$C]$_{\theta}$} & \colhead{[O$-$C]$_{\rho}$} & \colhead{Reference} \\         
\colhead{(2000)} & \colhead{Designation} & \colhead{name} & +2000 & & & \colhead{(deg)} & (mas) & ($''$) & (mas) & (mag) & \colhead{(deg)} & \colhead{($''$)}                   
& \colhead{code$^*$}  }   
\startdata     
 00143$-$2732 & HDS  33      & HIP  1144  & 12.9231 & y & 2 & 218.4 & 0.2 & 0.0671 &  0.3 & 1.2   &$-$78.9 & $-$0.001 & \citep{Cve2012}  \\ 
00417$-$2446 & B    10      & HD 2930    & 12.9231 & y & 2 & 164.5 & 0.8 & 0.0739 &  0.3 & 1.1   &        &          &           \\ 
00487$+$1841 & BU  495AB    & HIP  3795  & 13.7366 & I & 2 & 245.4 & 0.3 & 0.3454 &  0.3 & 0.6   &    1.7 & $-$0.023 & \citep{Sca2000b} \\ 
00522$-$2237 & STN   3AB    & HIP  4072  & 12.9231 & y & 2 & 242.7 & 0.3 & 1.9858 &  0.2 & 0.9 * &        &          &           \\ 
01061$-$4643 & SLR   1AB    & HIP  5165  & 13.7368 & y & 2 &  93.2 & 0.0 & 0.5093 &  0.1 & 0.5   &    0.8 & $-$0.012 & \citep{Alz2009a} \\ 
01071$-$0036 & HDS 144AB    & HIP  5245  & 13.7367 & I & 2 & 150.6 & 0.8 & 0.2893 &  2.4 & 2.2   &    0.2 &    0.000 & Table 7   \\ 
01071$-$0036 & BAG  12AC    & HIP  5245  & 13.7367 & I & 2 & 154.4 & 0.1 & 1.3642 &  5.6 & 4.7   &        &          &           \\ 
01144$-$0755 & WSI  70Aa,Ab & HIP  5799  & 12.9232 & y & 2 & 137.2 & 0.2 & 0.0219 &  2.3 & 1.3   &        &          &           \\
\enddata 
\end{deluxetable}   

\end{flushleft}

\begin{deluxetable}{l l l   c c c  c c   }                                                     
\tabletypesize{\tiny}                   
%\tabletypesize{\scriptsize}                   
%\rotate                                                                                        
\tablecaption{Unresolved stars                                                                  
\label{tab:single} }                                                                            
\tablewidth{0pt} 
 \tablenum{3}                                                                               
\tablehead{                                                                                     
WDS (2000) & \colhead{Discoverer} & \colhead{Hipparcos} & \colhead{Epoch} & \colhead{Filter} &  
\colhead{N} & \multicolumn{2}{c}{5$\sigma$ Detection Limit}          \\             
 & \colhead{Designation} & \colhead{or other} & \colhead{+2000} & & &                           
\colhead{$\Delta m (0\farcs15)$}  & \colhead{$\Delta m (1'')$}         \\             
 & \colhead{or other name}  & \colhead{name}  & & & & \colhead{(mag)} &  \colhead{(mag)}        
}                                                                                               
\startdata     
%00219$-$2300 &	RST5493BC &	HIP 1732 &	12.9231 &	y &	2 &	4.91 &	6.54  \\ 
%00490$+$1656 &	HIP 3810 &	HIP 3810 &	13.7367 &	I &	1 &	4.22 &	5.57  \\ 
%0            &	       &	           &	13.7367 &	y &	2 &	4.94 &	6.26  \\ 
%01144$-$0755 &	WSI 70Aa,Ab &	HIP 5799 &	12.9232 &	I &	1 &	3.88 &	6.12  \\ 
%01496$-$4646 &	TOK 184 &	HIP 8498 &	13.7368 &	y &	1 &	4.71 &	5.25  \\ 
%02128$-$0224 &	TOK 39Aa,Ab &	HIP 10305 &	12.9205 &	y &	2 &	4.47 &	6.88  \\ 
%02136$-$0849 &	TOK 361 &	HIP 10365 &	13.7367 &	y &	1 &	3.88 &	4.71  \\ 
%             &	       &	           &	13.7367 &	I &	2 &	3.12 &	4.59  \\ 
%02321$-$1515 &	TOK 382Aa,Ab &	HIP 11783 &	13.7478 &	y &	1 &	3.92 &	4.47  
% Discovery Designations changed by BDM:
%   HIP  10365 = TOK 361
%   HIP  11783 = TOK 382Aa,Ab
%   HIP  25148 = TOK  93Aa,Ab
%   HIP  26676 = RAS   1
%   HIP  27531 = TOK 363
%   HIP 105872 = TOK 372
%   HIP 108649 = TOK 215
%   HIP 114880 = TOK 217
%   HIP 116215 = TOK 219
%   HIP 117475 = TOK 376
%
00219$-$2300 &	RST5493BC &	HIP 1732 &	12.9231 &	y &	2 &	4.91 &	6.54  \\ 
00490$+$1656 &	HIP 3810 &	HIP 3810 &	13.7367 &	I &	1 &	4.22 &	5.57  \\ 
             &	       &	           &	13.7367 &	y &	2 &	4.94 &	6.26  \\ 
01144$-$0755 &	WSI 70Aa,Ab &	HIP 5799 &	12.9232 &	I &	1 &	3.88 &	6.12  \\ 
01496$-$4646 &	TOK 184 &	HIP 8498 &	13.7368 &	y &	1 &	4.71 &	5.25  \\ 
02128$-$0224 &	TOK 39Aa,Ab &	HIP 10305 &	12.9205 &	y &	2 &	4.47 &	6.88  \\ 
02136$-$0849 &	TOK 361 &	HIP 10365 &	13.7367 &	y &	1 &	3.88 &	4.71  \\ 
             &	       &	           &	13.7367 &	I &	2 &	3.12 &	4.59  \\
\enddata 
\end{deluxetable}   

\begin{deluxetable}{l l l  cc  r r l }                                                                                                                                
\tabletypesize{\scriptsize}                                                                                                                                                     
%\rotate                                                                                                                                                                         
\tablecaption{New  pairs
\label{tab:new} }                    
\tablewidth{0pt}  
\tablenum{4}                                                     
\tablehead{
\colhead{WDS} & \colhead{Discoverer} & \colhead{Other} & \colhead{Epoch}  &  \colhead{Filt} &\colhead{$\theta$} &                
\colhead{$\rho$}  & \colhead{$\Delta m$}  \\         
\colhead{(2000)} & \colhead{Designation} & \colhead{name} & +2000 & & \colhead{(deg)} &  ($''$) &  (mag) }                       
\startdata 
02321$-$1515 &	TOK\phn382Aa,Ab       & HIP 11783  & 13.7478 &	I & 341.8 &	0.2101 &        4.0 \\ 
04007$+$2023 &	TOK\phn363Aa,Ab       & HIP 18719  & 13.7369 &	y &  95.7 &	0.1023 &        5.2 \\ 
07435$+$0329 &	TOK\phn355Aa,Ab       & HIP 37645  & 13.1285 &	y & 330.2 &	0.2078 &        3.7 \\ 
08391$-$5557 &	TOK\phn356BC          & HIP 42424  & 13.1287 &	y & 309.4 &	0.0861 &        2.7 \\ 
09086$-$2550 &	TOK\phn357BC          & HIP 44874  & 13.1314 &	I & 219.2 &	0.1258 &        0.9 \\ 
09383$+$0150 &	TOK\phn358            & HIP 47292  & 13.1286 &	y & 344.4 &	0.4533 &        3.7 \\ 
11192$-$1950 &	TOK\phn383Aa,Ab       & HD 98412   & 13.2380 &	y &  25.8 &	0.0518 &        0.3 \\ 
11420$-$1701 &	TOK\phn384Aa,Ab       & HIP 57078  & 13.2381 &	y & 128.5 &	0.1467 &        3.6 \\
\enddata
\end{deluxetable}

\begin{deluxetable}{l l  cc c  }          
\tabletypesize{\scriptsize}     
\tablecaption{Likely spurious binaries
\label{tab:spurious} }                    
\tablewidth{0pt}                                                                                                           
\tablenum{5}
\tablehead{  
\colhead{WDS} & \colhead{Discoverer} & 
\colhead{Last} &
\colhead{SOAR} &
\colhead{$N_{\rm UR}$} \\
\colhead{(2000)} & \colhead{Designation} & 
\colhead{measure} &
\colhead{dates} &
  }
\startdata 
03182$-$6230 &	BNU\phn\phn\phn2Aa,Ab & 1978 & 2008--2012 & 3 \\
07383$-$2522 &  B\phn\phn\phn731      & 1966 & 2009--2013 & 5 \\
08246$-$0109 & B\phn\phn2527AB        & 1961 & 2010--2013 & 3 \\
08246$-$0345 & CHR\phn172Aa,Ab        & 1988 & 2011--2013 & 3 \\
08326$-$1502 & B\phn\phn2528          & 1942 & 2009--2013 & 4 \\
23114$-$4259 & B\phn\phn\phn594       & 1963 & 2008--2012 & 3 \\
\enddata
\end{deluxetable}

%\clearpage

\begin{deluxetable}{ll@{}r@{\,}lrrrrrrrc cc}
\tabletypesize{\scriptsize}
\tablenum{6}
\tablewidth{0pt}
\tablecaption{First Orbital Elements \label{tab:orbit-new} }
\tablehead{
\colhead{WDS} & 
\multicolumn{3}{c}{Discoverer} &
\colhead{~~~$P$} & 
\colhead{~~$T_\circ$} &
\colhead{~~$e$} & 
\colhead{~~$\Omega$} &
\colhead{~$a$} & 
\colhead{~~$i$} &
\colhead{~~$\omega$} &
\colhead{Gr.} &
\colhead{$\pi_{\rm HIP}$} &
\colhead{$\Sigma {\cal M}$}
 \\
\colhead{(Figure)} &
\multicolumn{3}{c}{Designation} & 
\colhead{~~~(yr)} & 
\colhead{~~(yr)} &
\colhead{ } & 
\colhead{~~($\circ$)} &
\colhead{~($''$)} & 
\colhead{~~($\circ$)} &
\colhead{~~($\circ$)} &
&
\colhead{(mas)} &  
\colhead{${\cal{M}}_{\odot}$}
}
\startdata
02022$-$2402 & TOK &   41 & Ba,Bb & 13.88 & 2006.3    & 0.09       & 134.4    &  0.099     & 154.2    & 232.1    & 3 & 16.1 & 1.2 \\
(3)          &     &      &   & $\pm$4.37 & $\pm$3.3  & $\pm$0.15  & $\pm$17  & $\pm$0.024 & $\pm$23  & $\pm$121 &     &    &  \\
02572$-$2458 & BEU  &4 &Ca,Cb & 1.519    & 2007.103   & 0.558 & 89.5 & 0.060 & 179.6 &  250.6  & 1 & 39.9 & 1.5 \\    
(4)          &     &      &   &     *     & $\pm$0.018& *          & $\pm$3.1 & $\pm$0.005 & *        & *        &     & &  \\
04008$+$0505 & A   & 1937 &       &       41.90        &           2014.1     &   0.699    &         185 &   0.096 &       41 &       38 & 2 & 5.4  &  3.2  \\   
(5)          &     &      &       &   $\pm$0.57        &     $\pm$27.0     & $\pm$0.038    &     $\pm$23 & $\pm$0.002 &       $\pm$143 &     $\pm$32 &   &                    \\
04215$-$2055 & B   & 1935 & AB & 21.22 & 2002.16 & 0.505 & 239.5 & 0.234 & 155.2 & 310.9&  5  & 25.9 & 1.6  \\
  (6)        &     &      &   & $\pm$0.19 & $\pm$0.22 & $\pm$0.028 & $\pm$19.1 & $\pm$0.006 & $\pm$4.8 & $\pm$20.0&     &  & \\
\enddata
\end{deluxetable}

\scriptsize
\begin{deluxetable}{ll@{}r@{\,}lrrrrrrrcl}
\tabletypesize{\scriptsize}
\tablenum{7}
\tablewidth{0pt}
\tablecaption{Revised Orbital Elements}
\tablehead{
\colhead{WDS} & 
\multicolumn{3}{c}{Discoverer} &
\colhead{~~~$P$} & 
\colhead{~~$T_\circ$} &
\colhead{~~$e$} & 
\colhead{~~$\Omega$} &
\colhead{~$a$} & 
\colhead{~$i$} &
\colhead{~~$\omega$} &
\colhead{Gr.} &
\colhead{Published Orbit} \\
 &
\multicolumn{3}{c}{Designation} & 
\colhead{~~~(yr)} & 
\colhead{~~(yr)} &
\colhead{ } & 
\colhead{~~($\circ$)} &
\colhead{~($''$)} & 
\colhead{~~($\circ$)} &
\colhead{~~($\circ$)} &
\colhead{ } & 
\colhead{Reference} 
}
\startdata
% WDS Disc P T e W a i w Gr 
01071$-$0036 & HDS &  144 & AB    &     \phn\phn72.6\phn\phn     &      2073.1\phn\phn   &     \phn0.237    &      \phn281.9 &     \phn0.365 &      \phn138.2\phn &     \phn\phn30.\phn & 4 & \citet{Cve2011a}  \\ % checked
%             &     &      &       &                  &                 &               &             &            &              &            &   &                   \\
04107$-$0452 & A   & 2801 &       &     \phn\phn20.42\phn       &           1952.61\phn    &  \phn0.887    &        \phn156.6 &  \phn0.194 &     \phn\phn71.5 &     \phn\phn70.4 & 2 & \citet{Baz1986a}     \\
             &     &      &       & \phn\phn$\pm$0.12\phn       & \phn\phn$\pm$0.36\phn    & $\pm$0.031    & \phn\phn$\pm$3.5 & $\pm$0.018 & \phn\phn$\pm$2.2 & \phn\phn$\pm$3.8 &   &                      \\
04258$+$1800 & COU & 2682 &       &     \phn\phn54.8\phn       &           1955.6\phn    &  \phn0.163    &        \phn328.5 &  \phn0.348 &     \phn\phn68.7 &     \phn\phn11.6 & 4 & \citet{Doc2007h}             \\
%             &     &      &       &                  &                 &               &             &            &              &            &   &                   \\
05059$-$1355 & A   & 3009 &       &        \phn284\phn       &           1957.2\phn    &  \phn0.237    &        \phn244.9 &  \phn1.283 &     \phn\phn40.1 &        \phn294.1 & 5 & \citet{Erc1985a}             \\
%             &     &      &       &                  &                 &               &             &            &              &            &   &                   \\
05135$+$0158 & STT &  517 & AB    &     \phn987.\phn\phn\phn     &      1925.3\phn\phn   &     \phn0.935    &      \phn138.\phn &     \phn0.77\phn &      \phn\phn18.\phn\phn &     \phn304.\phn & 4 & \citet{Msn2009}   \\
\enddata
\end{deluxetable}


\begin{thebibliography}{}


\bibitem[Alzner et al.\ (2009)]{Alz2009a} 
Alzner, A., Argyle, R.\ \& Anton, R.\ 2009, Inf.\ Circ.,   169

\bibitem[Argyle \& Alzner (2010)]{Ary2010} 
Argyle, R. W.\ \& Alzner, A.\ 2010, Inf.\ Circ.,   170

\bibitem[Balega et al. (1999)]{Bag1999b} 
Balega, I. I., Balega, Yu. Yu., Hofmann, K.-H. et al. 1999, SvAL, 25, 797
%Tokovinin, A. A.\ \& Weigelt, G.\ 

\bibitem[Balega et al. (2005)]{Bag2005} 
Balega, I. I., Balega, Y. Y., Hofmann, K.-H. et al.  2005, A\&A, 433, 591
%Pluzhnik, E.A., Schertl, D., Shkhagosheva, Z.U.\ \& Weigelt, G.\ 2005, A\&A 433, 591

\bibitem[Baize (1986)]{Baz1986a} 
Baize, P.\ 1986, A\&AS, 65, 551

\bibitem[Bender \& Simon (2008)]{Bender2008}
Bender, C. F. \& Simon, M. 2008, ApJ, 689, 416
%Astrophys. J., 689, 416-429 (2008)

\bibitem[Beuzit et al. (2004)]{Beuzit2004}
Beuzit, J.-L., Segransan, D., Forveille, T. et al. 2004, A\&A, 425, 997
%BEUZIT J.-L., SEGRANSAN D., FORVEILLE T., UDRY S., DELFOSSE X., MAYOR M., PERRIER C., HAINAUT M.-C., RODDIER C., RODDIER F. and MARTIN E.L. 

\bibitem[van den Bos (1961)]{B__1961d} 
van den Bos, W. H.\ 1961, Union Obs.\ Johannesburg Circ., 120, 380

\bibitem[Burnham (1913)]{Burnhan13} 
Burnham, S.W. 1913, Carnegie Inst. Wash.,  168

\bibitem[Cvetkovic et al. (2008)]{Cve2008a} 
Cvetkovic, Z., Novakovic, B.\ \& Todorovic, N.\ 2008, New Astronomy, 13, 125

\bibitem[Cvetkovic (2008)]{Cve2008c} 
Cvetkovic, Z.\ 2008, AJ, 136, 1746

\bibitem[Cvetkovic \& Ninkovic (2010)]{Cve2010e} 
Cvetkovic, Z.\ \& Ninkovic, S.\ 2010, AN, 331, 304

\bibitem[Cvetkovic (2011)]{Cve2011a} 
Cvetkovic, Z.\ 2011, AJ, 141, 116

\bibitem[Cvetkovic (2012)]{Cve2012} 
Cvetkovic, Z.\ 2012, Inf.\ Circ.,   177

\bibitem[Davis et al. (2005)]{Dvs2005} 
Davis, J., Mendez, A., Seneta, E. et al. 2005, MNRAS, 356, 1362
%Tango, W.J., Booth, A.J., O'Byrne, J.W., Thorvaldson, E., Ausseloos, M.\ et al.\ 2005, MNRAS 356, 1362

\bibitem[DeRosa et al. (2012)]{DRs2012} 
DeRosa, R. J., Patience, J., Vigan, A. et al. 2012, MNRAS, 422, 2765
%Wilson,  P.A.,  Schnieder,  A.,  McConnell, N.J.,  Wiktorowicz,  S.J.,
%Marois,  C., Song,  I., Macintosh,  B., Graham,  J.R.,  Bessell, M.S., Doyon, R.\ \& Lai, O. 

\bibitem[Docobo et al. (1994)]{Doc1994d}
Docobo, J., Ling, J.\ \& Prieto, C.\ 19994, ApJS, 91, 793

\bibitem[Docobo \& Tamazian (2007)]{Doc2007h}
Docobo, J. A.\ \& Tamazian, V. S.\ 2007, Inf.\ Circ.,  162

\bibitem[Docobo \& Ling (2008)]{Doc2008c} 
Docobo, J. A. \& Ling, J. F. 2008, Inf.\ Circ.,  164

%\bibitem[Docobo \& Ling (2008)]{Doc2009a} 
%Docobo, J. A. \& Ling, J. F. 2009, Inf.\ Circ.,  167

\bibitem[Docobo \& Ling (2009)]{Doc2009g} 
Docobo, J. A. \& Ling, J. F.\ 2009, AJ, 138, 1159

\bibitem[Docobo \& Ling (2010)]{Doc2010i} 
Docobo, J. A. \& Ling, J. F.\ 2010, Inf.\ Circ.,  172

\bibitem[Docobo \& Ling (2011)]{Doc2011d} 
Docobo, J. A. \& Ling, J. F.\ 2011, Inf.\ Circ.,  174

\bibitem[Docobo \& Campo (2011)]{Doc2011f} 
Docobo, J. A. \& Campo, P.\ 2011, Inf.\ Circ.,  174

\bibitem[Docoo \& Campo (2012)]{Doc2012a} 
Docobo, J. A. \& Campo, P.\ 2012, Inf.\ Circ.,  176

\bibitem[Docobo \& Andrade (2013)]{Doc2013d}
Docobo, J. A. \& Andrade, M.\ 2013, Inf.\ Circ.,  179

\bibitem[Dommanget (1979)]{Dom1979a}
Dommanget, J. 1979, Bull.\ R.\ Astron.\ Obs.\ Belgium, 9, 116


\bibitem[Ehrenreich et al. (2010)]{EHR10}
Ehrenreich, D., Lagrange, A.-M., Montagnier, G. et al. 2010, A\&A, 523, 73
%2010A&A...523A..73E - Astron. Astrophys., 523, A73-73 (2010) - 02.12.10 11.03.11 November(II) 2010
%EHRENREICH D.; LAGRANGE A.-M.; MONTAGNIER G.; CHAUVIN G.; GALLAND F.; BEUZIT J.-L.; RAMEAU J. 

\bibitem[Erceg (1985)]{Erc1985a}
Erceg, V. 1985, Bull.  Obs.  Astron. Belgrade,  135, 45
               
%\bibitem[ESA (1997)]{HIP}
%ESA 1997, The  Hipparcos and Tycho Catalogues, ESA SP-1200 (Nordwijk,
%Netherlands: ESA Publication Division) 

\bibitem[Finsen (1964)]{Fin1964b} 
Finsen, W. S. 1964, Republic Obs.  Circ.,  123, 59

\bibitem[Forveille et al. (1999)]{Frv1999} 
Forveille, T., Beuzit, J.-L., Delfosse, X. et al. 1999, A\&A, 351, 619
% Segransan, D.,  Beck, F.,  Mayor, M., Perrier,  C., Tokovinin,  A.  et

\bibitem[Griffin (2012)]{Griffin2012}
Griffin, R. F. 2012, JAA, 33, 29

\bibitem[Halbwachs et al. (2000)]{Halbwachs2000}
Halbwachs, J.-L., Arenou, F., Mayor, M. et al. 2000, A\&A, 305, 581

\bibitem[Hartkopf et al. (1996)]{Hrt1996a} 
Hartkopf, W. I., Mason, B. D.  \& McAlister, H. A.  1996, AJ, 111, 370

\bibitem[Hartkopf, McAlister \& Mason (2001)]{INT4} 
Hartkopf, W. I., McAlister, H. A. \ \& Mason, B .D.\ 2001, AJ 122, 3480

\bibitem[Hartkopf et al. (2000)]{Hrt2000a} 
Hartkopf, W. I., Mason, B. D., McAlister, H. A. et al. 2000, AJ, 119, 3084
%Roberts, L.C., Jr., Turner, N.H.,ten Brummelaar, T.A., Prieto, C.M., Ling, J. F.  \& Franz, O.G. 

\bibitem[Hartkopf (2000)]{Hrt2000b} 
Hartkopf, W. I.  2000, Inf.  Circ.,   141

\bibitem[Hartkopf \& Mason (2000)]{Hrt2000c} 
Hartkopf, W. I.  \& Mason, B. D.  2000, Inf.  Circ.,   142

\bibitem[Hartkopf, Mason \& Worley (2001)]{VB6} 
Hartkopf, W. I., Mason, B. D. \& Worley, C. E. 2001, AJ, 122, 3472 (VB6)
%(see the current version at 
%\url{http://www.usno.navy.mil/USNO/astrometry/ optical-IR-prod/wds/orb6.html})

\bibitem[Hartkopf \& Mason (2001a)]{Hrt2001a}
Hartkopf, W. I.  \& Mason, B. D.  2001a, Inf.  Circ.,   143

\bibitem[Hartkopf \& Mason (2001b)]{Hrt2001b} 
Hartkopf, W. I.  \& Mason, B. D.  2001b, Inf.  Circ.,   145

\bibitem[Hartkopf \& Mason (2010)]{Hrt2010a} 
Hartkopf, W. I.  \& Mason, B. D.  2010, Inf.  Circ.,   170

\bibitem[Hartkopf \& Mason (2011)]{Hrt2011d} 
Hartkopf, W. I.  \& Mason, B. D.  2011, Inf.  Circ.,   175

%\bibitem[Hartkopf, Tokovinin \& Mason (2012)]{HTM12}
%Hartkopf, W. I., Tokovinin, A., Mason, B. D. 2012, AJ, 143, 42 (HTM12)

\bibitem[Hartkopf et al. (2012)]{Hrt2012a} 
Hartkopf, W. I., Tokovinin, A.  \& Mason, B. D.  2012, AJ, 143, 42

\bibitem[Hartkopf \& Harshaw (2013a)]{Hrt2013a} 
Hartkopf, W. I.  \& Harshaw, R.  2013a, Inf.  Circ.,   179

\bibitem[Hartkopf \& Harshaw (2013b)]{Hrt2013c}
Hartkopf, W. I.  \& Harshaw, R.  2013b, Inf.  Circ. ,  181

\bibitem[Heintz (1978a)1978]{Hei1978a}
Heintz, W. D.  1978a, ApJS, 37, 71

\bibitem[Heintz (1978b)]{Hei1978c} 
Heintz, W. D. 1978b, ApJS, 37, 515

\bibitem[Heintz (1981)]{Hei1981a}
Heintz, W. D.  1981, ApJS, 45, 559

\bibitem[Heintz (1984)]{Hei1984b} 
Heintz, W. D.  1984, A\&AS, 56, 5

\bibitem[Heintz (1986a)]{Hei1986a}  
Heintz, W. D.  1986a, A\&AS, 64, 1

\bibitem[Heintz (1986b)]{Hei1986b} 
Heintz, W. D.  1986b, A\&AS, 65, 411

\bibitem[[Heintz (1988)]{Hei1988d} 
Heintz, W. D.  1988, A\&AS, 72, 543

\bibitem[Heintz (1990)]{Hei1990c} 
Heintz, W. D.  1990, A\&AS, 82, 65

\bibitem[Heintz (1991)]{Hei1991} 
Heintz, W. D.  1991, A\&AS, 90, 311

\bibitem[Heintz (1993)]{Hei1993d} 
Heintz, W. D.  1993, A\&AS, 98, 209

\bibitem[Heintz (1995)]{Hei1995} 
Heintz, W. D.  1995, A\&AS, 99, 693

\bibitem[Heintz (1996)]{Hei1996c} 
Heintz, W. D.  1996, AJ 111, 412

\bibitem[Heintz (1997)]{Hei1997}
Heintz, W. D.  1997, ApJS, 111, 335

\bibitem[Heintz \& Borgman (1984)]{Heintz84}
Heintz, W. D. \& Borgman, E. R. 1984, AJ, 89, 1068

\bibitem[Heintz (1984)]{Heintz}
Heintz, W. D. 1984, ApJ, 284, 806

\bibitem[Horch et al. (2002)]{Horch02}
Horch, E. P., Robinson, S. E., Meyer, R. D. et al. 2002, AJ, 123, 3442
% van Altena, W., Ninkov,  Z.\ \& Piterman, A.

\bibitem[Horch et al. (2006)]{Horch06} 
Horch, E. P. Davidson, J. W., Jr., van Altena, W. F. et al 2006, AJ, 131, 1000
%Girard, T.M.,  Lopez, C.E., Franz, O.G., \& Timothy, J.G. 


\bibitem[Horch et al. (2008)]{Horch08}
Horch, E. P., van Altena, W. F., Cyr, W. M., Jr. et al. 2008, AJ, 136, 312
%Kinsman-Smith, L.,  Srivastava, A.\ \& Zhou, J.\ 

\bibitem[Jasinta (1997)]{Jas1997}
Jasinta, D. M. D.  1997, Visual Double Stars, ed.  Docobo et al., 367

%\bibitem[Lang (1992)]{Lang}
%Lang, K. R. 1992, Astrophysical data. Planets and Stars (Berlin:
%Springer-Verlag)

\bibitem[Ling (2010)]{Lin2010c} 
Ling, J. F.  2010, AJ, 139, 1521

\bibitem[Makarov \& Kaplan (2005)]{MK05}
Makarov, V. V. \& Kaplan, G. H.,  2005, AJ, 129, 2420 (MK05)

\bibitem[Mante (2001)]{Mnt2001c}
Mante, R.  2001, Inf.  Circ.,   145

\bibitem[Mante (2003)]{Mnt2003b} 
Mante, R.  2003, Inf.  Circ.,   150

\bibitem[Mason \& Hartkopf (1999)]{Msn1999c} 
Mason, B. D.  \& Hartkopf, W. I.  1999, Inf.  Circ.,   138

\bibitem[Mason \& Hartkopf (2001)]{Msn2001c} 
Mason, B. D.  \& Hartkopf, W. I.  2001, Inf.  Circ.,   144

\bibitem[Mason et al.\ (2001)]{WDS}
Mason, B. D., Wycoff, G. L., Hartkopf, W. I. et al. 2001,  AJ, 122, 3466 (WDS)
%, Douglass, G. G. \& Worley, C. E. 2001a, 
%(see the current version at 
%\url{http://www.usno.navy.mil/USNO/astrometry/ optical-IR-prod/wds/wds.html})

\bibitem[Mason et al.  (2004)]{WSI2004a}
Mason, B. D., Hartkopf, W. I., Wycoff, G. L. et al. 2004, AJ, 127, 539
%Pascu, D., Urban, S.E., Hall, D.M., Hennessy, G.S., Rafferty, T.J.  et al.  

\bibitem[Mason et al. (2006)]{WSI2006b} 
Mason, B. D., Hartkopf, W. I., Wycoff, G. L.  \& Holdenried, E. R. 2006, AJ, 132, 2219

\bibitem[Mason \& Hartkopf (2005)]{Msn2005} 
Mason, B. D.  \& Hartkopf, W. I.  2005, Inf.  Circ.,   156

\bibitem[Mason et al. (2009)]{Msn2009}
Mason, B. D., Hartkopf, W. I., Gies, D. R. et al. 2009, AJ, 137, 3358
%Henry, T.J.  \& Helsel, J.W.  2009, AJ 137, 3358

\bibitem[Mason \& Hartkopf (2010)]{Msn2010a} 
Mason, B. D.  \& Hartkopf, W. I.  2010, Inf.  Circ.,   170

\bibitem[Mason et al. (2010)]{Msn2010c} 
Mason, B. D., Hartkopf, W. I.  \& Tokovinin, A.  2010, AJ, 140, 735

\bibitem[Mason \& Hartkopf (2011a)]{Msn2011a} 
Mason, B. D.  \& Hartkopf, W. I.  2011a, Inf.  Circ.,   173

\bibitem[Mason (2011)]{Msn2011b} 
Mason, B. D.  2011, Inf.  Circ.,   173

\bibitem[Mason \& Hartkopf (2011b)]{Msn2011c} 
Mason, B. D.  \& Hartkopf, W. I.  2011b, Inf.  Circ.,   174

\bibitem[Mason \& Hartkopf (2012)]{Msn2012a} 
Mason, B. D.  \& Hartkopf, W. I.  2012, Inf.  Circ.,   178

%\bibitem[Mason (2011)]{Mason11}
%Mason, B. D. 2011, private communication

\bibitem[Muller (1955)]{Mlr1955a} 
Muller, P.  1955, J.  Obs.,  38, 17

\bibitem[Muterspaugh et al. (2010)]{Mut2010b} 
Muterspaugh, M. W., Hartkopf, W. I., Lane. B. F. et al.  2010, AJ, 140, 1623
%O'Connell, J., Williamson, M., Kulkarni, S.R., Konacki, M., Burke, B.F., Colavita, M.M., Shao, M.  \& Wiktorowicz, S.J.  2010, AJ 140, 1623

\bibitem[Nordstr\"om et al. (2004)]{N04}
Nordstr\"om,  B., Mayor,  M.,  Andersen, J. et al. 2004, A\&A, 418, 989 
%Holmberg, J., Pont,  F., Jorgensen, B. R., Olsen, E.  H., Udry, S., \&
%Mowlavi, N.

\bibitem[Norro (1983)]{Nrr1983}
Norro, A. 1983, Bull. R. Astron. Obs. Belgium 9, 262

\bibitem[Olevic \& Jovanovic (2001)]{Ole2001} 
Olevic, D. \& Jovanovic, P.  2001, Serbian AJ, 163, 5

\bibitem[Olevic \& Cvetkovic (2004a)]{Ole2004b} 
Olevic, D.  \& Cvetkovic, Z.  2004a,  Inf.  Circ.,   152

\bibitem[Olevic \& Cvetkovic (2004b)]{Ole2004e} 
Olevic, D.  \& Cvetkovic, Z.  2004b, A\&A, 415, 259

\bibitem[Olevic \& Cvetkovic (2005)]{Ole2005d} 
Olevic, D.  \& Cvetkovic, Z.  2005, RevMexA\&A, 41, 17

\bibitem[Popovic (1969)]{Pop1969b} 
Popovic, G. M.  1969, Bull.  Obs.  Astron.  Belgrade, 27,  1, 33

\bibitem[Popovic (1978)]{Pop1978}
Popovic, G. M.  1978, Bull.  Obs.  Astron.  Belgrade,  129, 9

\bibitem[Raghavan et al. (2010)]{Raghavan10}
Raghavan, D., McAlister, H. A., Henry, T. J. et al. 2010, ApJS, 190, 1
% Latham, D. W., Marcy, G. W., Mason, B. D., Gies, D. R., White, R. J.,
% \& ten Brummelaar, Th. A.

\bibitem[Reffert \& Quirrenbach (2011)]{Reffert2011}
Reffert, S. \& Quirrenbach, A. 2011, A\&A, 527, 140

\bibitem[Riddle et al. (2014)]{RoboAO}
Riddle, R., Tokovinin, A., Mason, B. et al., 2014, ApJ, in preparation.

\bibitem[Rica Romero (2010)]{FMR2010c} 
Rica Romero, F. M.  2010, Inf.  Circ.,   171

\bibitem[Rica Romero \& Zirm (2011)]{FMR2011d} 
Rica Romero, F. M.  \& Zirm, H.  2011, Inf.  Circ.,   174

\bibitem[Rica (2012)]{FMR2012g} 
Rica, F. M.   2012, JDSO, 8, 127

\bibitem[Rica Romero (2012)]{FMR2012h} 
Rica Romero, F. M.  2012, Inf.  Circ.,   171

\bibitem[Rica Romero (2013)]{FMR2013e} 
Rica Romero, F. M.  2013, Inf.  Circ.,   181

\bibitem[Scardia (1983)]{Sca1983b}
Scardia, M.  1983, Inf.  Circ.,   90

\bibitem[Scardia et al. (2000)]{Sca2000b} 
Scardia, M., Prieur, J.-L., Aristidi, E.  \& Koechlin, L.  2000, AN, 321, 255

\bibitem[Scardia (2001)]{Sca2001b}
Scardia, M.  2001, Inf.  Circ.,   144

\bibitem[Scardia et al. (2007)]{Sca2007c} 
Scardia, M., Argyle, R. W., Prieur, J.-L. et al. 2007, AN, 328, 146
%Pansecchi, L., Basso, S., Law, N.M.  \& Mackay, C.D.  

\bibitem[Scardia et al. (2008)]{Sca2008d} 
Scardia, M., Prieur, J.-L., Pansecchi, L. \& Argyle, R. W. 2008, Inf.  Circ.,   165
 
\bibitem[Scardia et al. (2010)]{Sca2010c} 
Scardia, M., Prieur, J.-L., Pansecchi, L.  \& Argyle, R. W.  2010, Inf.  Circ.,   170

\bibitem[Seymour \& Hartkopf (1999)]{USN1999b} 
Seymour, D.  \& Hartkopf, W.  1999, Inf.  Circ.,   139

\bibitem[Seymour et al. (2002)]{USN2002} 
Seymour, D., Mason, B. D., Hartkopf, W. I.  \& Wycoff, G. L.  2002, AJ, 123, 1023

\bibitem[S\"{o}derhjelm (1999)]{Sod1999} 
S\"{o}derhjelm, S. 1999, A\&A, 341, 121

\bibitem[Starikova (1981)]{Sta1981a} 
Starikova, G. A.  1981, SvAL, 7, 130

\bibitem[Tokovinin (1999)]{Tok1999b} 
Tokovinin, A. A. 1999, SvAL, 25, 669

\bibitem[Tokovinin et al. (2005)]{Tok2005}
Tokovinin, A. A., Kiyaeva, O., Sterzik, M. et al.  2005, A\&A, 441, 695
%Orlov, V., Rubinov, A.  \& Zhuchkov, R.  2005, A\&A 441, 695

\bibitem[Tokovinin et al. (2008)]{SAM}
Tokovinin A., Tighe R., Schurter P. et al. 2008, Proc. SPIE, 7015, 157
%Cantarutti R., van der Bliek N.,
%Martinez M., Mondaca E., Montane A. 

\bibitem[Tokovinin \& Cantarutti (2008)]{TC08}
Tokovinin, A. \& Cantarutti, R. 2008, PASP, 120, 170 

\bibitem[Tokovinin (2012)]{Tok2012b} 
Tokovinin, A.  2012, AJ, 144, 56
%\bibitem[Tokovinin (2012)]{Fast}
%Tokovinin, A. 2012, AJ, 144, 56

\bibitem[Tokovinin (2013)]{Tok2013a} 
Tokovinin, A.  2013, AJ, 145, 76

\bibitem [Tokovinin, Mason \& Hartkopf (2010)]{TMH10}
Tokovinin, A., Mason, B. D., \& Hartkopf, W. I.  2010, AJ, 139, 743 (TMH10)

\bibitem[Tokovinin et al. (2010)]{SAM09}
Tokovinin, A. Cantarutti, R., Tighe, R. et al. 2010, PASP, 122,  1483 
%Schurter, P., van der Bliek, N., Martinez, M., Mondaca, E.


\bibitem[Tokovinin et al. (2012)]{astrom1} 
Tokovinin, A., Hartung, M., Hayward, Th. L., \& Makarov, V. V.  2012, AJ,
144, 7 

\bibitem[Tokovinin \& L\'epine (2012)]{LEP}
Tokovinin, A. \& L\'epine, S. 2012, AJ, 144, 102 

\bibitem[Tokovinin et al. (2013)]{astrom2} 
Tokovinin, A., Hartung, M., \&  Hayward, Th. L., 2013, AJ, 146, 8

\bibitem[Torres et al. (2006)]{SACY}
Torres, C. A. O., Quast, G. R., Da Silva, L. et al. 2006, A\&A, 460, 695 
% 2006A&A...460..695T
%TORRES C.A.O., QUAST G.R., DA SILVA L., DE LA REZA R., MELO C.H.F. and STERZIK M. 

\bibitem[Turner et al. (2008)]{Trn2008} 
Turner, N. H., ten Brummelaar, T. A., Roberts, L. C. et al. 2008, AJ, 136, 554
%Mason, B.D., Hartkopf, W. I.  \& Gies, D.R.  

%\bibitem[van Leeuwen (2007)] {HIP2}
%van Leeuwen, F. 2007, A\&A, 474, 653

\bibitem[Wilson (1976)]{WRH1976a} 
Wilson, R. H.  1976, MNRAS, 174, 75P

\bibitem[Zirm (2007)]{Zir2007} 
Zirm, H. 2007, Inf.  Circ.,   161

\bibitem[Zirm (2008)]{Zir2008} 
Zirm, H. 2008, Inf.  Circ.,   166

\bibitem[Zirm (2010a)]{Zir2010} 
Zirm, H. 2010a, Inf.  Circ.,   170

\bibitem[Zirm (2010b)]{Zir2010b} 
Zirm, H. 2010b, Inf.  Circ.,   172

\bibitem[Zirm (2013)]{Zir2013a} 
Zirm, H. 2013, Inf.  Circ.,   179


% \bibitem[ ()]{}
% \bibitem[ ()]{}
% \bibitem[ ()]{}
% \bibitem[ ()]{}
% \bibitem[ ()]{}


%

\end{thebibliography}
\end{document}